\documentclass[preprint,prb,showkeys]{revtex4}
\usepackage{graphicx}

\begin{document}

\title{Time-dependent density-functional theory for electronic excitations in materials: basics and perspectives}

\author{C. A.~Ullrich$^*$}
\affiliation{$\mbox{Department of Physics and Astronomy, University of Missouri, Columbia, MO 65211, USA}$.
\hspace*{5.5cm}$\mbox{Email: ullrichc@missouri.edu}$\hspace*{6cm}}

\author{V.~Turkowski}
\affiliation{$\mbox{Department of Physics and Astronomy, University of Missouri, Columbia, MO 65211, USA}$}
\affiliation{NanoScience Technology Center and Department of Physics, University of Central Florida, Orlando, Florida 32816, USA.
\hspace*{5cm}$\mbox{Email: turkowskiv@missouri.edu}$\hspace*{4.9cm}}

\date{\today}

\begin{abstract}
Time-dependent density-functional theory (TDDFT) is widely used to
describe electronic excitations in complex finite systems with large
numbers of atoms, such as biomolecules and nanocrystals. The first part of
this paper will give a simple and pedagogical explanation, using a two-level system, which shows how the
basic TDDFT formalism for excitation energies works.
There is currently an intense effort underway to develop TDDFT methodologies
for the charge and spin dynamics in extended systems, to calculate
optical properties of bulk and nanostructured materials, and to study
transport through molecular junctions. The second part of this paper highlights
some challenges and recent advances of TDDFT in these areas. Two examples are discussed:
excitonic effects in insulators and intersubband plasmon
excitations in doped semiconductor quantum wells.
\end{abstract}


\maketitle

\section{Introduction}

Many important areas of experimental and theoretical physics, chemistry, and materials science require an understanding
of the electronic excitations of atomic or molecular systems, nanostructures, mesoscopic systems, or bulk
materials.\cite{Carter2008,Koch2006,Scholes2006,Chelikowsky2003} A wide variety of spectroscopic techniques is being used
to characterize the electronic structure and
dynamics of these systems by probing their excitation spectra. The performance of any nanoelectronic device, such
as a molecular junction, is dominated by its electronic excitations. \cite{Nitzan2003,Koentopp2008}

Time-dependent density-functional theory (TDDFT) \cite{TDDFT_book} is an increasingly popular, universal
approach to electronic dynamics and excitations. Just like ground-state density-functional theory (DFT),
which is based on a set of rigorous theorems \cite{HK,KS} proving a one-to-one correspondence between
ground-state densities and potentials, there is a similar existence theorem for TDDFT, due to Runge and Gross, \cite{Runge1984}
which establishes the time-dependent density as a fundamental variable.

The usage of TDDFT as a practical tool to calculate excitation energies started in the mid-90's with the
groundbreaking work of Petersilka {\em et al.} \cite{Petersilka1996} and Casida. \cite{Casida1996}
In the Casida-formalism, the excitation energies are obtained from the following eigenvalue problem:
\begin{equation} \label{casida}
\left( \begin{array}{cc} {\bf A} & {\bf K} \\ {\bf K}^* & {\bf A}^* \end{array}\right)
\left(\begin{array}{c} {\bf X} \\ {\bf Y}\end{array}\right)
=
\omega
\left( \begin{array}{cc} -{\bf 1} & {\bf 0} \\ {\bf 0} & {\bf 1} \end{array}\right)
\left(\begin{array}{c} {\bf X} \\ {\bf Y}\end{array}\right),
\end{equation}
where the matrices $\bf A$ and $\bf K$ are defined as follows:
\begin{equation}
A_{ia\sigma,i'a'\sigma'} = \delta_{ii'}\delta_{aa'}\delta_{\sigma\sigma'}(\varepsilon_{a\sigma}-\varepsilon_{i\sigma})
+K_{ia\sigma,i'a'\sigma'}
\end{equation}
and
\begin{equation} \label{Kmat}
K_{ia\sigma,i'a'\sigma'} =\int d^3r \int d^3r' \varphi^*_{i\sigma}({\bf r}) \varphi_{a\sigma}({\bf r})
\left[ \frac{1}{|{\bf r}-{\bf r}'|} + f_{\rm xc,\sigma \sigma'}({\bf r},{\bf r}',\omega)\right]
\varphi_{i'\sigma'}({\bf r}') \varphi^*_{a'\sigma'}({\bf r}').
\end{equation}
Here, $\varphi_{i\sigma}$ and $\varepsilon_{i\sigma}$ are the Kohn-Sham orbitals and eigenvalues coming from
a self-consistent ground-state DFT calculation; we use the standard convention that $i,i',\ldots$ are indices of
occupied orbitals and $a,a',\ldots$ refer to unoccupied orbitals. $f_{\rm xc,\sigma\sigma'}({\bf r},{\bf r}',\omega)$
is the so-called exchange-correlation (XC) kernel,\cite{Gross1985} which is in general a frequency-dependent quantity, but in practice
is often approximated using frequency-independent expressions. This is known as the adiabatic approximation.

The eigenvalues $\omega$ of Equation (\ref{casida}) are,
in principle, the {\em exact} excitation energies of the system, provided one would start from an exact Kohn-Sham ground-state
calculation and then use the exact $f_{\rm xc,\sigma\sigma'}$. In practice, of course, static and dynamical XC functionals
need to be approximated. The formalism of Equations (\ref{casida})-(\ref{Kmat}) can also be recast in the shape of
an eigenvalue problem for the squares of the excitation energies:
\begin{equation}\label{eigen}
\sum_{a'i'\sigma'} \left[ \delta_{ii'}\delta_{aa'}\delta_{\sigma\sigma'}\omega_{ia\sigma}^2
+ 2\sqrt{\omega_{ia\sigma}\omega_{i'a'\sigma'}} K_{ia\sigma,i'a'\sigma'}
\right]\xi_{a'i'\sigma'}=\omega^2 \xi_{ai\sigma}\:,
\end{equation}
where $\omega_{ia\sigma} = \varepsilon_{a\sigma}-\varepsilon_{i\sigma}$.

This TDDFT formalism for excitation energies has become very popular for practical applications, and is nowadays implemented as an integral
part of several widely used software packages in theoretical chemistry. \cite{Furche2002} Overall, TDDFT affords a
unique balance between accuracy and efficiency, allowing the user to study systems that would be impossible
to treat with traditional wavefunction methods, for example in all-electron studies of the photochemistry of large biomolecules.
\cite{Marques2005,Varsano2006} To describe localized and delocalized electronic excitations in large conjugated and aggregated
molecules, TDDFT has also been used along semiempirical Hamiltonian models or approaches based on the Kohn-Sham density matrix
such as the collective electronic oscillator (CEO) model. \cite{Tretiak2002,Berman2003}

The broad spectrum of applications of TDDFT for excited states has been recently reviewed by Elliott {\em et al.} \cite{Elliott2007}
From the large body of available literature, the following trends for molecules have emerged. Transition frequencies, calculated with
standard gradient-corrected XC functionals are typically good to within 0.4 eV. Excited-states structural properties
such as bond lengths, vibrational frequencies and dipole moments are essentially as good as those of ground-state
calculations (about 1 \% for bond lengths, and about 5 \% for dipoles and vibrational frequencies).
Most importantly for large systems, the computational costs scale as $N^3$, versus $N^5$ for wavefunction methods of
comparable accuracy (eg CCSD, CASSCF).

While there exist efficient iterative algorithms for solving the full eigenvalue problem (\ref{casida}), it is
nevertheless useful to consider approximations, since these can lead to further insight.
The Tamm-Dancoff approximation (TDA) neglects the off-diagonal matrices $\bf K$ in Equation (\ref{casida}),
which results in the simpler eigenvalue problem
\begin{equation}\label{TDA}
{\bf AX}=\omega_{\rm TDA}{\bf X} \:.
\end{equation}
Physically, the TDA is the TDDFT analog of the configuration interaction singles (CIS) method.\cite{Hirata1999}
The TDA has some technical advantages away from ground-state equilibrium geometry, as discussed by Casida
{\em et al.} in Ref. \onlinecite{TDDFT_book}.

In an even more drastic approximation, Equation (\ref{eigen}) is truncated down to a $1\times 1$ matrix. This yields
the so-called small-matrix approximation (SMA),  \cite{Appel2003} which for spin-saturated systems is given by
\begin{equation}\label{SMA}
\omega^2_{\rm SMA} = \omega_{ia}^2 + 4\omega_{ia}K_{ia,ia}.
\end{equation}
This can be further approximated if the correction to the bare Kohn-Sham excitation energy $\omega_{ia\sigma}$ is small,
which is known as single-pole approximation (SPA):\cite{Petersilka1996,Appel2003}
\begin{equation}\label{SPA}
\omega_{\rm SPA} = \omega_{ia} + 2K_{ia,ia}.
\end{equation}
This approximation can also be viewed as a TDA for a two-level system. Below, we shall
present an alternative, more direct derivation of the SMA and SPA.
It turns out that these two approximations perform surprisingly well for simple closed-shell atomic systems. \cite{Vasiliev1999}
This might perhaps be viewed as merely a curiosity --- after all, one can describe such systems
with the full Casida TDDFT formalism without resorting to any approximation.
However, the SMA and SPA can prove very useful for situations such as extended  systems
where the full Casida formalism cannot be easily applied, and we shall hence focus on such cases.

The purpose of this paper is to present a discussion of simplified TDDFT approaches to
excitation energies in bulk materials and quantum wells. We will begin with a simple and pedagogical derivation of the small-matrix
and single-pole approximation for a two-level Kohn-Sham system. We will then show how these expressions
can be easily modified and generalized for the case of collective excitations in extended systems. In this way, one can arrive
at a simple treatment of excitonic effects in bulk insulators, and plasmon-like excitations
in doped semiconductor nanostructures. Some examples will be discussed.

\section{Excitation energies of a two-level Kohn-Sham system}
Let us consider a two-level system consisting of two orbitals $\varphi_1({\bf r})$ and $\varphi_2({\bf r})$ which
are eigenstates of the static Kohn-Sham Hamiltonian
\begin{equation}
H^0 = -\frac{\hbar^2 \nabla^2}{2m} + V_{\rm ext}({\bf r}) + \int d^3r \frac{n_0({\bf r})}{|{\bf r} - {\bf r}'|}
+ V_{\rm xc}[n_0]({\bf r})\:,
\end{equation}
where $V_{\rm xc}$ is the exchange-correlation (xc) potential, a functional of the ground-state density $n_0({\bf r})$.
We assume that $\varphi_1$ is doubly occupied and $\varphi_2$ is empty.

Now consider a weak perturbation $\lambda H'({\bf r},t)$ acting on the system. According to time-dependent perturbation
theory, the time evolution of the system is given by
\begin{equation}
\varphi({\bf r},t) = c_1(t) \varphi_1({\bf r}) + \lambda c_2(t) \varphi_2({\bf r}) \:.
\end{equation}
Let us construct the density matrix of the system as follows:
\begin{equation}
\rho(t) =
\left( \begin{array}{cc} \rho_{11} & \lambda \rho_{12} \\ \lambda \rho_{21} & \lambda^2 \rho_{22} \end{array}\right)
=
\left( \begin{array}{cc} |c_1|^2 & \lambda c_1^* c_2 \\ \lambda c_1 c_2^* & \lambda ^2 |c_2|^2 \end{array}\right),
\end{equation}
where we explicitly indicate the order of the perturbation through orders of $\lambda$. The density matrix
obeys the following equation of motion:
\begin{equation}
\dot{\rho} = -i[H,\rho] = -i [H^0 + \lambda H'(t),\rho].
\end{equation}
Dropping terms of order $\lambda^2$, this yields the time evolution of the
off-diagonal elements of the density matrix as follows:
\begin{eqnarray}
\dot{\rho}_{12} &=& -i[(H^0_{11}-H^0_{22})\rho_{12} -  H'_{12}\rho_{11}] \\
\dot{\rho}_{21} &=& i[(H^0_{11}-H^0_{22})\rho_{21} -  H'_{21}\rho_{11}],
\end{eqnarray}
where $H^0_{11}=\int d^3r \varphi_1({\bf r}) H^0 \varphi_1({\bf r})$ and similar for all other matrix elements of $H^0$ and $H'$.
Since $\rho_{11} = 1 + {\cal O}(\lambda^2)$, and defining $H^0_{22}-H^0_{11}=\omega_{21}$ (the bare Kohn-Sham excitation energy),
this simplifies to
\begin{eqnarray}
\dot{\rho}_{12} &=& i[\omega_{21}\rho_{12} +  H'_{12}] \\
\dot{\rho}_{21} &=& -i[\omega_{21}\rho_{21} +  H'_{21}].
\end{eqnarray}
Next, we make the ansatz (which will be justified later)
\begin{equation} \label{RWA}
\rho_{12}(t) = \tilde{\rho}_{12}(\omega)e^{-i\omega t} + \tilde{\rho}_{12}(-\omega)e^{i\omega t} \:,
\end{equation}
and similar for $\rho_{21}$, $H'_{12}$, and $H'_{21}$. This gives
\begin{eqnarray} \label{a1}
-\omega \tilde{\rho}_{12}(\omega) &=& [\omega_{21}\tilde{\rho}_{12}(\omega) +  \tilde{H}'_{12}(\omega)]\\
-\omega \tilde{\rho}_{21}(\omega) &=& -[\omega_{21}\tilde{\rho}_{21}(\omega) +  \tilde{H}'_{21}(\omega)] \label{a2}
\end{eqnarray}
and an additional two equation for $\tilde{\rho}_{12}(-\omega)$ and $\tilde{\rho}_{21}(-\omega)$
which do not contain any new information. Adding Equations (\ref{a1}) and (\ref{a2}) gives
\begin{equation} \label{e3}
\tilde{\rho}_{12}(\omega) + \tilde{\rho}_{21}(\omega) =
-\frac{\tilde{H}'_{12}(\omega)}{\omega_{21}+ \omega} - \frac{\tilde{H}'_{21}(\omega)}{\omega_{21}- \omega} \:.
\end{equation}
Let us now consider the perturbing Hamiltonian:
\begin{equation}
H'({\bf r},\omega) = \int d^3r' \left[\frac{1}{|{\bf r}-{\bf r}'|} +
f_{\rm xc}({\bf r},{\bf r}',\omega)\right] \delta n({\bf r}',\omega) \:,
\end{equation}
where
\begin{equation}
\delta n({\bf r},\omega)= 2 \varphi_1({\bf r}) \varphi_2({\bf r})[\tilde{\rho}_{12}(\omega)+ \tilde{\rho}_{21}(\omega)].
\end{equation}
Notice that we do not consider an external perturbation, only the linearized Hartree and XC potentials.
We are thus looking for an ``eigenmode'' of the system in a steady-state. This justifies the
ansatz (\ref{RWA}) made above.
We define the double matrix element
\begin{equation}
F(\omega) = 2\int d^3r \int d^3r' \varphi_1({\bf r}) \varphi_2({\bf r}) \left[\frac{1}{|{\bf r}-{\bf r}'|} +
f_{\rm xc}({\bf r},{\bf r}',\omega)\right]\varphi_1({\bf r}') \varphi_2({\bf r}') \: ,
\end{equation}
and Equation (\ref{e3}) becomes
\begin{equation}\label{poles}
\tilde{\rho}_{12}(\omega) + \tilde{\rho}_{21}(\omega) =
-  F(\omega)[\tilde{\rho}_{12}(\omega) + \tilde{\rho}_{21}(\omega)]
\left[\frac{1}{\omega_{21}+ \omega} + \frac{1}{\omega_{21}- \omega}\right].
\end{equation}
Canceling $\tilde{\rho}_{12}+\tilde{\rho}_{21}$ on both sides results in
\begin{equation}
1 = -\frac{2\omega_{21}}{\omega_{21}^2 - \omega^2} \: F(\omega)
\end{equation}
which gives the final result
\begin{equation}\label{small}
\omega^2 = \omega_{12}^2 + 2\omega_{21} F(\omega) \:.
\end{equation}
This is the same as the SMA, Equation (\ref{SMA}). From the point of view of our two-level system,
our derivation shows that the SMA considers the excitation $1\to 2$ as well as the de-excitation $2\to 1$. The
SPA, Equation (\ref{SPA}), on the other hand, only includes the excitation $1\to2$ [it is obtained by
ignoring the first pole in Equation (\ref{poles})]. In general,
the TDA (\ref{TDA}) ignores all de-excitations.

\section{Optical excitations in bulk insulators}

The standard ab-initio treatment of excitation processes in
insulators and semiconductors, including correlation-induced
screening, is based on many-body Green's function techniques such as
GW and the Bethe-Salpeter equation. \cite{Onida2002}
However, TDDFT has recently emerged as an alternative, computationally convenient
approach to electronic excitation processes in materials.
\cite{Onida2002,Reining,Kim,Botti2007} As discussed above,
excitation energies can be calculated in principle exactly in TDDFT, provided
the XC kernel $f_{\rm xc}({\bf r},{\bf r}',\omega)$ is known, and one starts with an exact ground-state calculation.
This statement is true not only for finite systems such as molecules, but remains valid
for extended systems. In Refs. \onlinecite{Onida2002,Reining}, an approximate $f_{\rm xc}$ was
constructed from many-body Green's functions, whereas Ref.
\onlinecite{Kim} uses an exact-exchange approach, including a cutoff in
wavevector space which mimics screening of the Coulomb interaction.
\cite{Bruneval2006} These studies have established that TDDFT is
capable of describing excitonic effects in solids, although one has
to use XC functionals that go beyond the more common ones such as
the adiabatic local-density approximation. \cite{TDDFT_book} The
resulting agreement with experimental data is excellent,
\cite{Botti2007} but the technical effort is not significantly less
than that of standard many-body approaches.

\begin{figure}
\centering
\includegraphics[origin=b,angle=0,width=0.4\linewidth]{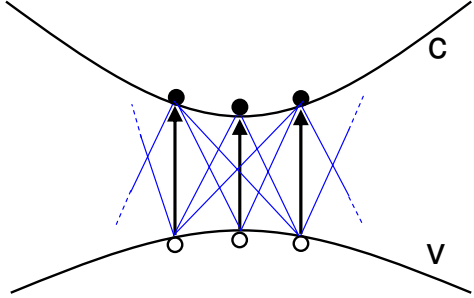}
\caption{Schematic illustration of interband transitions in an insulator. The thick arrows represent single-particle
transitions between the valence (v) and the conduction (c) band. Via dynamical many-body effects, these transitions
are coupled (thin blue lines). Excitons can thus be viewed as collective excitations, where the electron-hole interactions
involve the entire bands. } \label{fig1}
\end{figure}

 In the following,
let us for simplicity consider a two-band insulator with valence and conduction band energies $\varepsilon_{\bf k}^v$ and
$\varepsilon_{\bf k}^c$ and associated Bloch functions $\psi_{v{\bf q}}({\bf r})$ and  $\psi_{c{\bf q}}({\bf r})$.
The single-particle interband transition energies are simply given by $\omega_{\bf k}^{cv}=\varepsilon_{\bf k}^c-\varepsilon_{\bf k}^v$.
A TDDFT formalism for nonlinear ultrafast interband excitations has recently been developed, \cite{Turkowski2008}
which describes the electron dynamics of a two-band system via a generalization of the semiconductor Bloch equations.
This work has shown that TDDFT is capable of producing excitonic signatures in the interband dynamics.
We now briefly outline a simpler and more direct TDDFT approach for excitonic binding energies.\cite{Turkowski2009}

To describe excitonic effects in the interband absorption spectrum of insulators, one needs to go beyond
the single-particle picture and include dynamical many-body effects. In TDDFT, this is accomplished through
the dynamical XC kernel $f_{\rm xc}({\bf r},{\bf r}',\omega)$.
Let us illustrate this for the simple case of
our two-band insulator. The SPA of Equation (\ref{SPA}) refers to two discrete levels, but it can be
generalized to two entire bands as follows:
\begin{equation} \label{singlepole}
\sum_{\bf q} \left[\omega_{\bf q}^{cv}\delta_{\bf kq} + F_{\bf kq}(\omega)\right]\zeta_{\bf q}(\omega)
=\omega \zeta_{\bf k}(\omega) \:,
\end{equation}
where
\begin{equation} \label{fxcmat}
F_{\bf kq}(\omega) = \frac{2}{\Omega^2} \int_\Omega d^3r \!\int_\Omega
d^3r'\: \psi_{c{\bf k}}^*({\bf r})\psi_{v{\bf k}}({\bf r})f_{\rm
xc}({\bf r},{\bf r}',\omega)
\psi_{v{\bf q}}^*({\bf r}')\psi_{c{\bf q}}({\bf r}') \:.
\end{equation}
Thus, in contrast to the simple SPA Equation (\ref{SPA}) which gives an explicit expression for the
excitation energy, Equation (\ref{singlepole}) represents an eigenvalue equation which couples interband
transitions of different wavevectors via the coupling matrix $F_{\bf kq}$, Equation (\ref{fxcmat}).
This shows explicitly that excitons are collective excitations which involve not only the states at
the valence-band maximum and conduction-band minimum (for direct excitons), but the states of the entire bands.
A schematic illustration is given in Figure \ref{fig1}. Explicit evaluations of Equations (\ref{singlepole})
and (\ref{fxcmat}) are currently in progress.\cite{Turkowski2009}

The TDDFT formalism of Ref. \onlinecite{Turkowski2008} can in principle also be used to treat higher-order correlation effects such
as biexciton formation. These effects are especially important in the description of the ultrafast nonlinear optical response
in semiconductors (bulk and nanostructured). \cite{Chemla2001} Such nonlinear effects, which are entirely governed by dynamical correlations,
can be expected to put severe demands on the XC functionals, and will be the subject of future studies.

\section{Intersubband plasmons in quantum wells}

\begin{figure}
\centering
\includegraphics[origin=b,angle=0,width=0.4\linewidth]{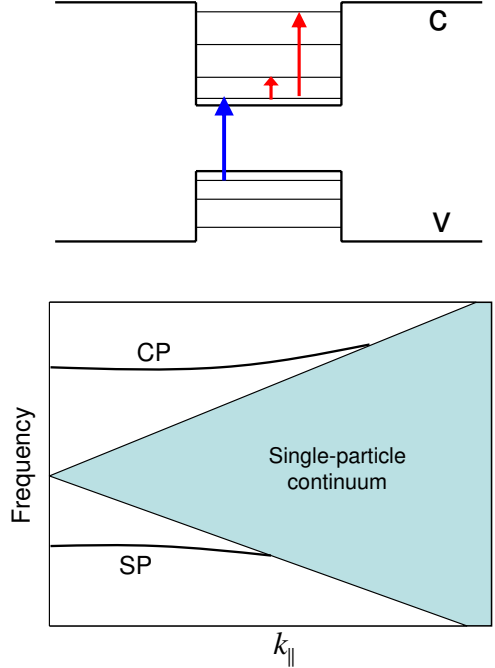}
\caption{Top: Schematic illustration of interband transitions (blue arrow) and intersubband transitions (red arrows) in a square quantum well.
Bottom: Frequency versus in-plane wave vector $k_{||}$ of intersubband excitations in a quantum well. Collective charge and spin plasmon
excitations (CP and SP) are separated from the region of incoherent single-particle excitations.} \label{fig2}
\end{figure}

The terahertz frequency regime is scientifically rich, but despite recent progress it is
technologically still underdeveloped. \cite{ISBbooks}
Subband spacings in typical III-V quantum wells are of the order of a few tens of meV.
Since a photon energy of 10 meV corresponds to a frequency of 2.4 THz, intersubband  transitions
in quantum wells appear as natural candidates for device applications to fill the ``Terahertz gap''
in the electromagnetic spectrum. A schematic illustration of intersubband transitions is given in the top panel
of figure \ref{fig2}.

The fundamental coupling mechanism between electromagnetic waves and carriers in a doped semiconductor
quantum well is through a collective excitation, the so-called intersubband plasmon.
To date, most applications of TDDFT linear response theory in quantum wells have been concerned with calculating
intersubband plasmon dispersions and linewidths. \cite{bloss89,jogai91,ryan91,dobson92,marmorkos93,ullrich98,ullrich01,ullrich02}
Due to its highly collective nature, and since the dynamics is essentially one-dimensional,
intersubband plasmons in quantum wells are an ideal testing ground for new TDDFT approaches.
A review of recent applications of TDDFT to the electron dynamics in semiconductor nanostructure is given
in Ref. \onlinecite{ullrich2006}.

Let us now extend our discussion to include charge as well as spin plasmon excitations.
The physical picture of these excitations is that in a charge plasmon the spin-up and spin-down electrons in the quantum well move
collectively in phase, whereas in a spin plasmon they move 180 degrees out of phase.
These excitations have a dependence on the in-plane wavevector ${\bf k}_{||}$, which is
schematically illustrated in Fig. \ref{fig2}. In the $\omega - k_{||}$ plane, one distinguishes an area of incoherent
single-particle excitations (also known as Landau damping), and two discrete plasmon branches. The charge plasmon lies above
the single-particle regime, and the spin plasmon lies below. Experimentally, this is confirmed via inelastic light
scattering spectroscopy.\cite{Pinczuk1989}

We shall in the following discuss how TDDFT describes the charge and spin plasmons in quantum wells.
We adopt the effective-mass approximation for electronic states in doped semiconductor nanostructures. Let
$N_s$ be the electronic sheet density (the number of electrons per area), which is chosen so that only
the lowest subband is occupied. It is assumed that the direction of growth, and thus of quantum confinement,
of the quantum well is along the $z$-axis, and electrons are free to move in the $x$-$y$ plane.
For simplicity, we consider only the first and the second subband.

The intersubband charge and spin plasmon frequencies $\omega_{c,s}$ are determined by the condition that
the following $2\times 2$ matrix equation,
\begin{equation}\label{eigen1}
\sum_{\sigma'} M_{\sigma\sigma'}({\bf k}_{||},\omega) h_{\sigma'}({\bf k}_{||},\omega) = \lambda h_{\sigma}({\bf k}_{||},\omega) \:,
\end{equation}
has eigenvalue $\lambda=1$. The matrix elements are given by
\begin{eqnarray} \label{matrixelements}
M_{\sigma\sigma'}({\bf q},\omega)
&=&
F_{21}({\bf q},\omega)   \left\{\int \! dz\!\int \!dz'
\left[\frac{2\pi}{q}\: e^{-q|z-z'|} +f^{\rm ALDA}_{{\rm xc},\sigma \sigma'}(z,z')\right]
\varphi_1(z) \varphi_2(z)\varphi_1(z') \varphi_2(z') \right.\nonumber\\
&-&
\frac{i\omega}{\omega_{21}^2}\int dz \left[
\left(\zeta_{{\rm xc},\sigma\sigma'}(z,\omega) + \frac{4}{3} \eta_{{\rm xc},\sigma\sigma'}(z,\omega)\right)
\left( \nabla_z \frac{P(z)}{n(z)}\right)^2 \right.\nonumber\\
&&\left.
{}+ \left.q^2 \eta_{{\rm xc},\sigma\sigma'}(z,\omega)\ \left(\frac{P(z)}{n(z)}\right)^2
-\frac{\sigma\sigma'}{4}\:\Re \rho_{\uparrow \downarrow}(z,\omega)\: P_{z}^2(z)\right] \right\},
\end{eqnarray}
where
\begin{equation}
P(z) = \varphi_1(z) \nabla_z \varphi_2(z) -  \varphi_2(z) \nabla_z \varphi_1(z) \:,
\end{equation}
and
\begin{equation}\label{ffun}
F_{21}({\bf q},\omega) =
-\int \frac{d^2k}{(2\pi)^2} \left\{ \frac{f(\varepsilon_{1}+k^2/2)}
{\omega +\omega_{21} -{\bf q}\cdot {\bf k} + \frac{q^2}{2} + i\delta}
+
 \frac{f(\varepsilon_{1}+k^2/2)}
{-\omega +\omega_{21} -{\bf q}\cdot {\bf k} + \frac{q^2}{2} - i\delta}
\right\}.
\end{equation}
Here, $\varphi_1(z)$ and $\varphi_2(z)$ are the envelope functions of the first two electronic subbands,
$\varepsilon_{1,2}$ are the associated Kohn-Sham energy eigenvalues, and $\omega_{21} = \varepsilon_2 - \varepsilon_1$
is the bare subband spacing (assuming parabolic subbands). A detailed derivation of Equations (\ref{matrixelements})-(\ref{ffun})
will be published elsewhere.\cite{Damico2009}

The matrix elements $M_{\sigma\sigma'}$ of Equation (\ref{matrixelements}) contain various XC contributions.
The first XC term features $f_{\rm xc,\sigma\sigma'}^{\rm ALDA}(z,z')$ in the adiabatic local-density approximation (ALDA). \cite{TDDFT_book}
This term is frequency independent and real, and therefore by itself gives rise to plasmon excitations that have no imaginary part,
and thus have an infinite lifetime.

In reality, since quantum wells are extended systems, plasmons are intrinsically damped.
Finite plasmon linewidths arise from the other XC terms in Equation (\ref{matrixelements}), featuring the
viscosity coefficients $\eta_{\rm xc,\sigma\sigma'}$ and $\zeta_{\rm xc,\sigma\sigma'}$ and the spin Coulomb drag
transresistivity $\rho_{\uparrow \downarrow}$. These terms originate \cite{VUC,Qianvignale1,Qianvignale2} in the
context of time-dependent current-DFT, and have been extensively studied for quantum wells.\cite{ullrich01,ullrich02,ullrich2006,Damico2006}

One can simplify things for the case of vanishing in-plane wavevector, $k_{||}=0$, and ignoring the XC contributions beyond ALDA
(which in general have only a minor influence on the plasmon frequency).
In that case, one obtains a straightforward generalization of the SMA for quantum wells as follows:
\begin{equation}\label{3.15}
\omega^2= \omega_{21}^2 + 2 \omega_{21}
(S_{\uparrow\uparrow} \pm S_{\uparrow\downarrow}) \:,
\end{equation}
where
\begin{equation}\label{3.14}
S_{\sigma\sigma'} = \frac{N_s}{2} \int \!dz \!\int\!dz'
\Big[-2\pi |z-z'| + f_{\rm xc,\sigma\sigma'}^{\rm ALDA}(z,z')\Big]
\varphi_{1}(z) \varphi_{2}(z) \varphi_{1}(z') \varphi_{2}(z')\:.
\end{equation}
The Hartree contribution in $S_{\sigma\sigma'}$ is known as ``depolarization shift'',
and the XC contribution is sometimes (somewhat misleadingly) called ``excitonic shift''.
The Hartree part always induces an up-shift of the plasmon frequency with respect to
$\omega_{21}$, and the XC part gives a smaller down-shift. In the charge plasmon,
the positive shift dominates, but for the spin plasmon,
the Hartree parts cancel out in Eq. (\ref{3.15}) and the spin plasmon frequency is redshifted with respect
to the single-particle excitation region, see Fig. \ref{fig2}. This is a remarkable result: the existence of the collective
intersubband spin plasmon is purely a consequence of XC effects.

The picture that emerges from these studies is that TDDFT has been extremely successful in describing excitations in
doped semiconductor nanostructures (which are essentially metallic systems, in contrast with the insulators
discussed in the previous section), already at the level of the simplest XC approximations. Currently, work is
in progress where the formalism is generalized to more realistic descriptions of the semiconductor materials,
for example including spin-orbit coupling effects. \cite{ullrichflatte1,ullrichflatte2,Damico2009}

\section{Conclusion and challenges for TDDFT}

In this paper, we have given a brief overview of the TDDFT methodology to calculate excitation energies in materials.
The basic formalism, which is widely used in popular quantum chemistry software packages, is based on Casida's equation
(\ref{casida}). This approach is universal and in principle capable of producing the exact excitation energies
of any system and material, provided that XC effects are properly accounted for.

Besides the full formalism, it is instructive to consider simplified approaches which are based, in one way or
another, on truncations of the matrices $\bf A$ and $\bf K$ of Equation (\ref{casida}), leading to the SMA or SPA.
We have shown how these approaches can be formulated for excitations in bulk materials and doped semiconductor
quantum wells. This allows for a straightforward discussion of excitonic effects in insulators, and intersubband
plasmons in quantum wells. In both cases we have given explicit expressions for calculating the collective
excitations. For extended systems, TDDFT not only gives the excitation energies in principle exactly, but
also the linewidths.

However, the success of TDDFT critically depends upon the quality of the XC functionals used. Simple local and
semilocal functionals, such as the LDA or GGA, usually work fine for most applications for molecular systems.\cite{Elliott2007}
On the other hand, there are situations which are difficult to deal with, namely a proper description of
double excitations or of charge-transfer excitations in large molecules. In these cases, one needs to go
beyond the adiabatic approximation and use an XC kernel $f_{\rm xc}({\bf r},{\bf r}',\omega)$ which is explicitly
frequency-dependent.\cite{Maitra1,Maitra2}
The exploration of non-adiabatic XC effects in TDDFT remains a subject of intense investigation.\cite{ullrich2006b}

For extended insulating or semiconducting systems (inorganic as well as organic),
the situation is even more complicated. A proper description of excitonic effects in
the optical absorption of such materials requires long-range XC kernels.\cite{Onida2002}
Standard XC functionals such as LDA and GGA fail to produce any excitonic effects, and are in fact
completely incapable of shifting the Kohn-Sham band edge.\cite{Gruning2007,Izmaylov2008}
It has been demonstrated \cite{Reining,Kim,Botti2007,Bruneval2006} that one can construct XC kernels which
produce absorption spectra in excellent agreement with experiment. However, these kernels are technically involved
and do not lead to major computational advantages compared to Green's function based techniques.
It remains an important task for TDDFT to construct parameter-free expressions for $f_{\rm xc}$ which accurately reproduce excitonic effects
in solids, yet are simple enough to implement in practice. Work along these lines is in progress.

\acknowledgments
This work was supported by NSF Grant DMR-0553485.

\end{document}